\documentclass{PoS}
\usepackage{afterpage}
\title{Overview of the lithium problem in metal-poor stars and new results
on $^6$Li}
\ShortTitle{Overview of the lithium problem}
\author{\speaker{R.~Cayrel$^a$}, M.~Steffen$^b$, 
P. Bonifacio$^{acd}$, H.-G. Ludwig$^{ac}$ and E.~Caffau$^a$\\
\llap{$^a$}GEPI, Observatoire de Paris, CNRS, Universit\'e Paris Diderot,61 av. 
de l'Observatoire 75014 Paris, France\\
\llap{$^b$} Astrophysikalisches Institut Potsdam, An der Sternwarte 16, Potsdam, Germany\\
\llap{$^c$} CIFIST Marie Curie Excellence Team\\
\llap{$^d$}Istituto Nazionale di Astrofisica,
Osservatorio Astronomico di Trieste, Via G.B. Tiepolo 11, 34143 Trieste, Italy\\
E-mail: \email{roger.cayrel@obspm.fr}}
\abstract{Two problems are discussed here. The first one is the 0.4~dex 
 discrepancy between the $^7$Li abundance derived from the spectra of
 metal-poor halo stars on the one hand, and from Big Bang nucleosynthesis,
 based on the cosmological parameters constrained by the Wilkinson Microwave 
 Anisotropy Probe (WMAP) measurements, on the other hand. Lithium, indeed, can 
 be depleted in the convection zone of unevolved  stars, by a combination of 
 diffusion  and slow mixing with the hotter layers below the convection zone, 
 where $^7$Li is destroyed by the $^7$Li(p,$\alpha)^4$He reaction. 
 The understanding of the hydrodynamics of this crucial zone near the bottom of
 the convective envelope in dwarfs or turn-off stars of solar
 metallicity  has recently made enormous progress with the inclusion of
 internal gravity  waves. However, similar work for metal-poor stars is still
 lacking. So in spite of several investigations claiming to have partly explained 
 the above mentioned $^7$Li discrepancy, it is not yet clear whether or not the 
 depletion occurring in the metal-poor stars themselves is adequate to produce a 
 $^7$Li plateau. The second problem concerns  the difficulties met in accounting
 for the large amount of $^6$Li recently found in metal-poor halo stars 
 (Asplund et al. 2006 \cite{ALN06}). It has already been suggested 
 (Cayrel et al. 2007 \cite{CSC07}) that the convection-related asymmetry of 
 the $^7$Li line could mimic the signal attributed so far to the weak blend of 
 $^6$Li in the red wing of the $^7$Li line. However, this suggestion was only
 based on a hydrodynamical simulation for a single set of atmospheric parameters,
 representing the halo turn-off star HD~74000. Now  the theoretical line asymmetry 
 has been computed for an extended range in effective temperature, gravity and 
 metallicity, covering the stars of the Asplund et al.\ sample.
The computed asymmetry is about two per cent, a value which is also the mean of
the $^6$Li/$^7$Li ratio of the full sample determined with symmetric profiles by
Asplund et al. 2006. Our conclusion is that these    
observations  can be reinterpreted in terms of intrinsic line asymmetry, plus 
the amount  of $^6$Li/$^7$Li  given by Asplund et al. reduced by 0.02. This
drastically reduces the number of certain $^6$Li detections, from 9 to 1 or 2.}
     
     \FullConference{10th Symposium on Nuclei in the Cosmos\\
		 July 27 - August 1 2008\\
		 Mackinac Island, Michigan, USA}
\begin{document}
\section{The WMAP / Spite plateau discrepancy}
The Big Bang nucleosynthesis leads now to very accurate predictions of the
primordial abundances, with the cosmological parameter 
$\eta$=$N$(baryons)/$N$(photons) constrained by the observations of WMAP. On the 
usual scale of abundances by number of atoms per 10$^{12}$ hydrogen atoms, the
logarithmic abundance of $^7$Li is $\log n$(Li)$\,=\,$$2.64 \pm 0.04$ (Spergel 
et al. 2007 \cite{WMAP07}), against a logaritmic abundance of $2.22 \pm 0.08$ for 
the Spite plateau (Charbonnel \& Primas 2005 \cite{CP05}). The first idea which
comes to mind is that the nuclear fragility of $^7$Li can deplete the element 
by the same mechanism which has depleted this nuclide in the Sun by about 2 dex
in 4.6 Gyr. The main difficulty with this scheme is that the Spite plateau is 
both flat and shows only a very small scatter. It is {\it a priori} difficult to 
imagine a mechanism leading to a constant depletion over two decades in metallicity 
and almost 1000~K in effective temperature. The problem of the small scatter was 
also serious as long as effects of stellar rotation were believed to be a main 
ingredient of the depletion mechanism. The inital distribution of stellar rotation 
induces without doubt a statistical scatter. We shall come back to this mechanism 
in subsection \ref{S:mixing}.\\
Other possible mechanisms have been proposed. In section \ref{S:astration}  
we shortly discuss the proposal that the depletion of the Spite plateau is 
due to a massive astration of the cosmic matter before the birth of the galactic
halo stars (Piau et al.\ 2006 \cite{PBB06}). In section \ref{S:particles} we quote 
proposals using supersymmetric particles, very briefly, because the subject is 
considered elsewhere in this volume.
In section \ref{S:mixing} we discuss in detail the diffusion -- slow mixing --
nuclear burning mechanism, now widely studied in dwarfs of solar metallicity.  

\subsection{A major astration of cosmic matter before the formation of halo
stars}
\label{S:astration}
Piau et al.\ 2006 \cite{PBB06}
have proposed that about half of the mass of the cosmic matter from which the 
halo stars formed was already processed by pregalactic supernovae (SNe) of 
masses between $10$ and $40$~$M_{\odot}$. Although this scenario has some
interesting features concerning, for example, the high depletion of Li in the 
extremely iron poor, C-enriched  star HE 1327-2326, the level of astration 
invoked seems incompatible with the expected resulting metallicity on the 
part of the plateau below [Fe/H]=-2.0 (Prantzos 2006, NIC-IX \cite{P07}).

\subsection{Supersymmetric particles}
\label{S:particles}
If they exist, supersymmetric particles may have played a role, both in
depleting  $^7$Li and in producing $^6$Li during the Big Bang, winning on 
the two sides. As the topic has been developed elsewhere in this volume, 
I will just mention here a few references dealing with this topic:
Jedamzik 2004 \cite{JED04}, Pospelov 2007 \cite{POS07}, Cumberbatch et 
al.\ 2007 \cite{CUM07}, and Kusakabe et al.\ 2007 \cite{KUS07}.  

\subsection{Depletion by the diffusion -- slow mixing -- nuclear burning
mechanism}
\label{S:mixing}
Atmospheres of low mass unevolved stars are rapidly mixed by convection and
a small amount of convective  overshooting. Below the convection zone, the 
situation is rather complex. Heavy ions settle down very slowly, but if a new 
ionization stage is reached, radiative support by continuous plus line absorption 
can block this motion. Under the impulse of Georges Michaud (see the Meeting in 
Honor of Georges Michaud 2005 \cite{ARV05}) an enormous amount of work has been 
done in this field. The situation is complicated by the existence of hydrodynamical 
effects which tends to modify this vertical stratification and by the existence
of meridional circulation. Initially, turbulence generated by differential 
rotation was considered to be the main cause of mixing, inhibiting to some degree 
the effects of diffusive sedimentation. 
The presence of hydrodynamical turbulence which tends to scramble the effects 
of diffusion, also enables the depletion of fragile nuclides which are destroyed 
at higher temperatures some depth below the bottom of the convection zone. 
In that case, the hydrodynamical turbulence creates a slow mixing
between the convection zone and the burning region below. Very clearly, this is 
the case for Sun, in which the photospheric $^7$Li is depleted  by more than a
factor $100$ from an ISM initial value of $3.3$. 
Korn et al.\ 2007 \cite{KGR07} have constrained the role of diffusion by comparing 
the abundances of a few elements in unevolved stars and evolved stars in the
globular cluster NGC~6397. They derive a depletion of $^7$Li by about $0.25$~dex
using a stellar model of O. Richard (see \cite{ARV05}). This value depends upon the
difference of effective temperature between the unevolved stars and the evolved
stars, and upon the empirical model of Richard, done before the theoretical 
 work of Charbonnel \& Talon 2008 \cite{CT08}. 
The latter approach has been very successful in explaining a large body of 
observations able to constrain the theory. First the 
evolution with time can be checked by observing stars in clusters of various
ages, from the Pleiades to M67. Then the Boesgaard-Tripicco dip \cite{BT86}
poses a severe challenge to the theory, which was resolved only
very recently \cite{CT08}.  The simultaneous observation of the three 
fragile nuclides $^7$Li, $^7$Be, and $^9$B in field stars (Boeesgaard et 
al.\ 1998 \cite{BDS98}) supplies direct
evidence that the depletion is linked to the physical properties of the nuclei,
and reaches levels unattainable by diffusion only. This body of observations has
allowed to put the theory of loss of angular momentum and of chemical mixing
in cool stars on a much better footing than before. A key ingredient has been
to take into consideration internal gravity waves as well as the effects of
rotation (meridional circulation and shear turbulence) \cite{CT08}.
Unfortunately the low-metallicity stars of the galactic halo
have not yet been investigated in as much detail as population~I stars.
Further work is needed before a well accepted conclusion can be drawn.        

\subsection{New observational results on the metal-poor tail of the Spite
plateau}
\label{S:newobs}
\begin{figure}
\centering
\includegraphics[width=0.8\textwidth,clip=true]{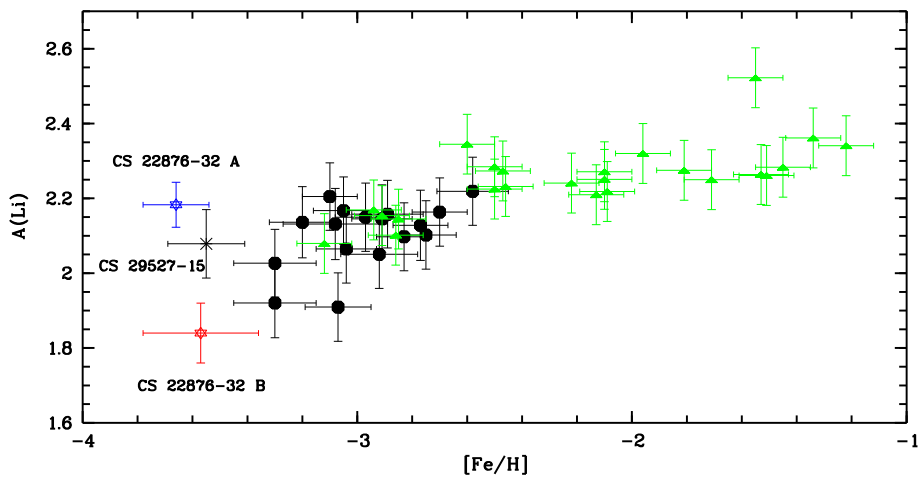}
\caption{The metal-poor end of the Spite plateau. Note the increased
scatter, and the trend towards a drop of the Li abundance at the lowest
mean metallicities. Adapted from Bonifacio et al.\ 2007 \cite{BMS07} and
Gonz\'alez Hern\'andez et al.\ 2008 \cite{GHB08}.}
\label{endplateau}
\end{figure}

Very few unevolved stars at extremely low metallicity are bright enough to
be subjected to the spectroscopic study of the weak Li feature. A sudden drop in 
the $^7$Li abundance at metallicities below 0.01 the solar metallicity is quite 
visible in a diagram plotted not in logarithmic coordinates, but on a linear 
scale (see Fig.~22 in Asplund et al. 2006 \cite{ALN06}, referred to as ALN06 
further on).
In a study of the unevolved metal-poor halo stars, Bonifacio et al.\ 2007 
\cite{BMS07} have studied 17 objects with metallicities between 
$-3.8 <$ [Fe/H] $< -2.7$ (see Fig.~\ref{endplateau}).
This study supports the conclusion that in this metallicity range something 
is happening to the Spite plateau:
(i) the scatter in Li abundance is much larger here than on the 
plateau ($-2.5 <$ [Fe/H] $< -1.5$). The double-lined spectroscopic binary 
CS~22876-032 is a striking example of this behavior. Component A
(higher mass) has  a logarithmic $^7$Li abundance of 2.2,
 component B  has a Li abundance of only 1.75 (\cite{GHB08}, see 
also Fig.~\ref{endplateau}).
(ii) a decline of the abundance of $^7$Li at very low metallicity
looks probable.  Garcia P\'{e}rez 2008 \cite{Perez08} also finds a drop at 
these lowest metallicities, which can be connected to the strict upper 
limit of $^7$Li < $0.6$ found in the ultra-metal-poor star
HE 1327-2326 at [Fe/H]= -5.3 (Frebel et al. 2008 \cite{FCE08}).

\section{ The $^6$Li problem}
\label{S:6Li}
 
Until 2006, only very few determinations of the abundance of\ $^6$Li were performed.
The observation is very difficult, the resonance line of\ $^6$Li being a weak 
doublet, blended with the corresponding doublet of\ $^7$Li, shifted by only 
0.16 \AA\ to the red of the\ $^7$Li line, and about 20 times weaker. The 
doublets are not  resolved because of the thermal and turbulent broadening of the
lines. 
So, the presence of\ $^6$Li is detected only by a slight extra depression of the red 
wing of the unresolved  $^7$Li feature (Fig.~\ref{profile1}).
A drastic change occurred with the paper by Asplund et al. 2006 \cite{ALN06}, who
published a set of 24 determinations of\ $^6$Li abundances, based on high quality 
spectra acquired with ESO's VLT/UVES combination. The surprising result was a
kind of\ $^6$Li plateau (top of Fig.\,23 of their paper) at the level 
$\log n$($^6$Li) $\approx$ 0.8. As clearly demonstrated by their Figure 23 
this behavior does not fit the expectations based on cosmic ray (CR) production 
(again their Fig.~23, bottom). This result has immediately triggered new 
theoretical papers trying to explain the new\  $^6$Li plateau.

\subsection{ The pregalactic Cosmic Ray solution}
Rollinde, Vangioni \& Olive 2006 \cite{RVO06} have considered pregalactic cosmic
rays generated by shocks around population~III objects. By collisions between
accelerated and at rest helium nuclei, the so called $\alpha$
fusion process generates both Lithium isotopes by  the reactions
$^4$He($\alpha$, D)$^6$Li and  $^4$He($\alpha$, p)$^7$Li.
This process works, but needs a very large fraction of the kinetic energy 
of population~III SNe to be converted into CR energy, as noted by Prantzos 2006
\cite{P06} and Evoli et al.\ 2008 \cite{ESF08}. It seems also difficult to 
produce a plateau -- in the diagram $\log n$($^6$Li) vs. [Fe/H] -- in the   
hierarchical model of Galaxy formation of Evoli et al.\ 2008.
Other scenarios involving CR production have been proposed, but are exposed 
to similar critics.

\subsection{The supersymmetric particle solution}  
See the list of references given in section \ref{S:particles}, and paper
by B. Fields (NIC-X, this volume).

\subsection{ The $^7$Li asymmetry solution: new results}
\label{S:asymmetry}
Observationally, the $^6$Li measurement relies on the detection of a slight 
extra depression of the red wing of the $^7$Li feature. However,
such a line asymmetry already exists without the presence of $^6$Li. 
3D hydrodynamical models of convection produced by the CO$^5$BOLD code 
\cite{WFSL04} clearly predict a convection-induced line asymmetry of similar
strength. Cayrel et al.\ 2007 \cite{CSC07} have shown that, in the case of 
the star HD74000, a halo turn-off star of metallicity [Fe/H] = -2.0, 
the derived $^6$Li abundance is reduced to a negligible  amount when
taking into account this intrinsic line asymmetry. It was also shown that 
iron lines with equivalent widths similar to those of the doublet components 
of the $^7$Li feature, and with the same stratification, show about the same 
amount of asymmetry as the computed one for\ $^7$Li. This is the only 
``sanity'' check suggesting that our computed asymmetries are correct. The referee 
asked us if we could compare our asymmetries with those obtained with 
the 3D models used in ALN06, for example. This was not possible for $^7$Li,
by lack of  published data. We tried to use the published data in Allende Prieto
et al. \cite{AAGL02} on iron bisectors for Procyon. However in the short time 
available we do not claim to have reached a definite conclusion. Work is in 
progress to investigate this important point.    
  
The big question is to know what amount of\ $^6$Li would remain in the 
Asplund et al.\ analysis of 24 halo stars if the natural asymmetry of\
$^7$Li had been properly included. In order to get some idea of that, 
we have extended the computation of the theoretical asymmetry of the\ 
$^7$Li feature to other values of the parameters metallicity, effective 
temperature and gravity, spanning the range of Asplund et al.\ observations. 
Specifically, new computations have been performed for the central effective 
temperature of the sample, $T_{\rm eff}$$\,=\,$$6300$~K, at the 3 metallicities 
[Fe/H]=$-3.0$, $-2.0$, and $-1.0$, and for the two gravities $\log g$\,=\,$4.0$ 
(turn-off) and $4.5$ (unevolved dwarfs). 
At the central metalliciy, [Fe/H]=$-2.0$, the computations were made also for
$T_{\rm eff}$$\,=\,$$5900$~K and again for  $\log g$\,=\,$4.0$ and $4.5$.  
The results of these computations are presented below.
  
\begin{figure}
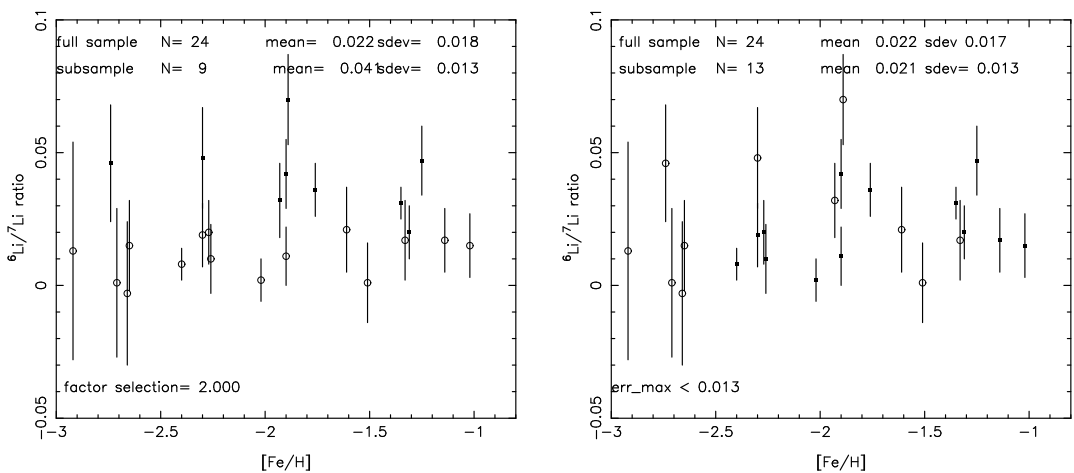

\centering
\includegraphics[width=.45\textwidth]{fig2B.eps}
\hspace{1em}
\includegraphics[width=.45\textwidth]{fig3B.eps}
\caption{left panel:Fig.~13 (middle) of ALN06. Selected objects: filled dots,
discarded objects: open circles. The mean value of $n$($^6$Li)/$n$($^7$Li)
of the full sample is $0.022$, the mean value of the selected subsample 
is 0.041. Right panel: Same as left panel excepted tha the selection
is now done on the accuracy only error $< 0.013$. The mean value is not affected
now}
\label{bias1}
\end{figure}

\subsubsection{A statistical problem with the subsample with detected $^6$Li}
    
Before proceeding further, it is worth to note that what everybody calls the\
$^6$Li plateau  is the almost constant $^6$Li abundance of the stars having a
certain detection of\  $^6$Li, namely those satisfying the condition:
$(^6\mathrm{ Li})/(^7\mathrm{Li})  > 2\sigma$ ; 
 $\sigma$ being the estimated accuracy 
in the determination of this ratio. As the abundance of\ $^7$Li is almost constant 
in the sample, and the noise is also fairly constant, it means that all values 
of $n$( $^6$Li)/$n$($^7$Li) below about 2 per cent have been discarded. 
This is clear in  Fig.~13 of ALN06, reproduced here as Fig.~\ref{bias1}.  
We are interested in the meaning of the ''plateau''  
 with respect to the full sample we note that the mean of the ratio for the 
full sample  
is $0.022$, compared to $0.041$ for the  subsample.
The median of the full sample is 0.018. A selection criterion involving
the value of the studied variable is well known to introduce a severe bias
(see the warning of Kapteyn \& Weersma 1910 \cite{KW10} about small parallaxes).   
A selection with $n$($^6$Li)/$n$($^7$Li) $ > 3\sigma$ would move the plateau 
 to a still higher value of $0.045$, whereas a selection on the error
only, keeping errors below $0.013$, when the mean error is $0.0154$, leads to 
a mean of $n$($^6$Li)/$n$($^7$Li)=0.021, almost unchanged (Fig.~\ref{bias1}).
A more severe constraint on the accuracy of the determinations, error $<$ 0.012,
gives a mean of 0.18, still close to the mean of the full sample.
In what follows we claim that the true measured signal is about 2 per cent 
induced by the asymmetry of\ $^6$Li (affected by the 
measurement error). The mean of 0.022 acquires then a new signification:
absence of\  $^6$Li. The certain detections are now decreased in number, as 
only values of $2\sigma$ above 0.02 represent certain detections. The new number is 
of the order of 1 or 2, and critically depends on the value of $\sigma$.   

\begin{figure}
\centering
 \includegraphics[width=.6\textwidth]{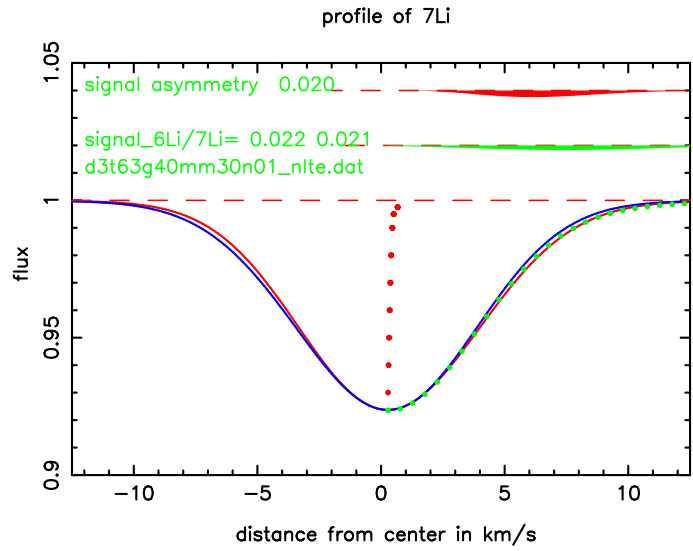}
 \caption{The red line is a (intrinsically asymmetric) 3D-NLTE profile 
 of the major component of $^7$Li (area $W$), computed for atmospheric 
 parameters $T_{\rm eff}=6250$~K, $\log g=4.0$, [Fe/H] $=-3.0$. The blue 
 line  is the mirror image (with respect
 to the  bisector of the line core) of the red line completing a symmetric 
 profile with the left wing of the red profile. The difference between the right
 wings of the red and the blue profiles (area $\Delta W$) measures the amount of 
 asymmetry of the 3D profile (signal $\equiv \Delta W/W = 0.020$). The 
 green dots represent the profile of a $^6$Li / $^7$Li blend, computed with the 
 symmetric profile and a ratio $n$($^6$Li)/$n$($^7$Li)=0.022, equal to the mean 
 value of the observations of ALN06 for a selection of stars with parameters 
 $6100$~K$<T_{\rm eff}<6400$~K and $3.7<\log g<4.3$.
 The signal of this blend measures the isotopic ratio:  
 $n$($^6$Li)/$n$($^7$Li)$\,=\,$$\Delta W/W$$\,=\,$$0.021$.
 There is a slight difference between the signals produced by the asymmetric
 3D profile and the $^6$Li / $^7$Li blend (the $^6$Li blend is slightly more 
 red-shifted and shallower), but these differences are too small
 to be measurable.}
 \label{profile1} 
\end{figure}
\begin{figure}
\centering
\includegraphics[width=.6\textwidth]{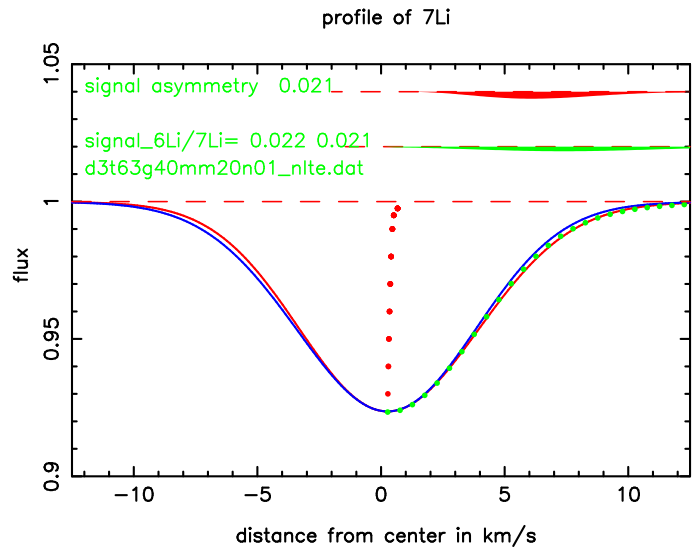}
\caption{Same as fig~\protect\ref{profile1}, except that here [Fe/H]= -2.0.}
\label{profile2}
\end{figure} 

\begin{figure}
\centering
\includegraphics[width=0.5\textwidth,angle=90]{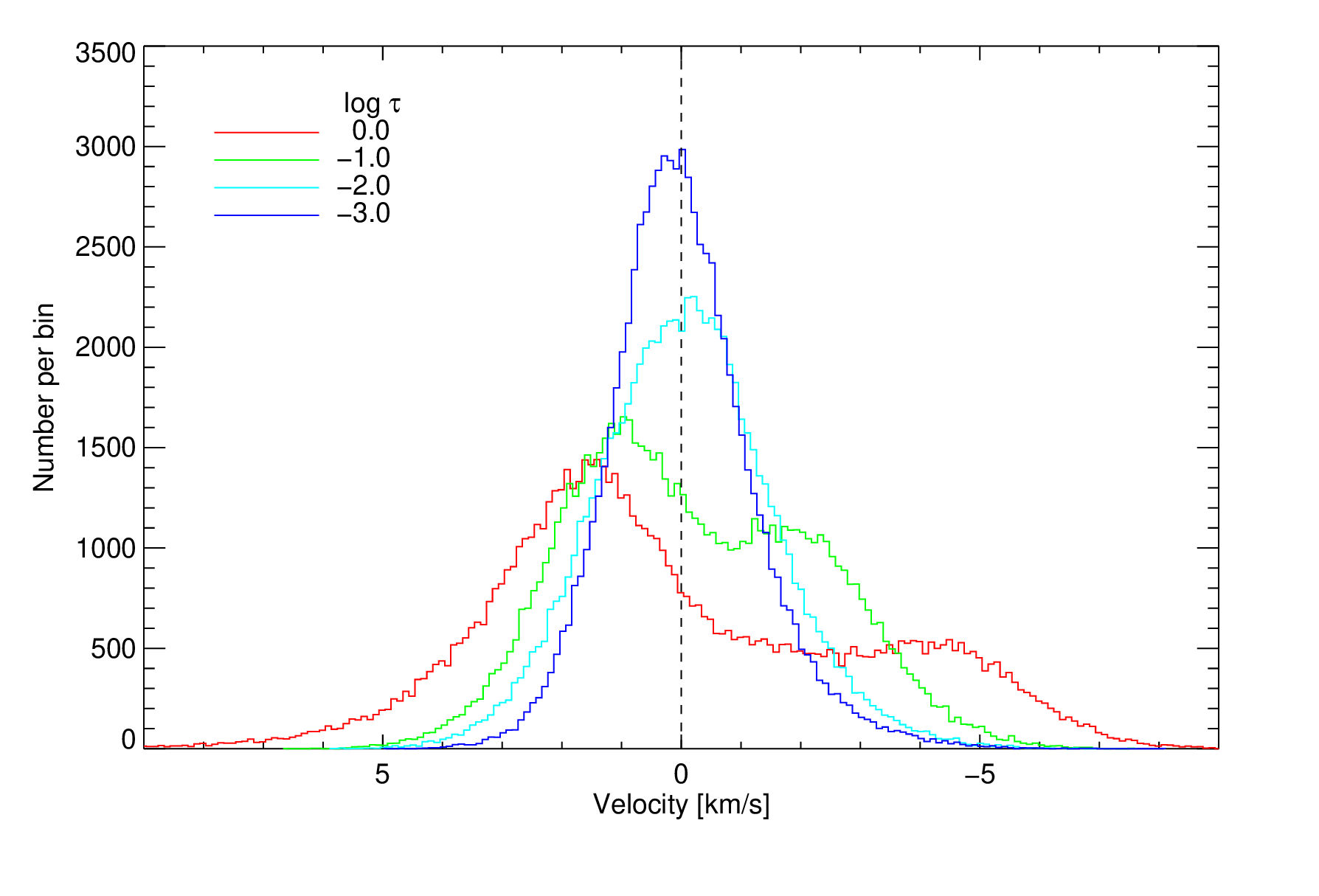}
\caption{Histogram of vertical velocities at 4 optical depths in the atmosphere of
a typical metal-poor star ($T_{\rm eff}=6250$~K, $\log g=4.0$, [Fe/H]$=-2.0$). 
Positive (negative) velocities correspond to blue- (red-) shifts. See text for details}
\label{z-motion}
\end{figure}

\subsubsection{The size and behavior of the convective line asymmetry} 
Our main interest is investigate the dependence of the convective line 
asymmetry upon stellar metallicity, effective temperature and surface
gravity.
Table~\ref{assymprofile1} gives our preliminary results for a small 
grid of 8 hydrodynamical simulations.
The entry `asym' is the excess of absorption in the red wing of the blend 
relative to the equivalent width of the full line. 
'Asym' was determined by fitting the asymmetric 3D line profile, obtained
from NLTE line formation calculations on a number ($N$) of selected 
snapshots from the hydrodynamical simulations, with a combination of two 
symmetric 1D line profiles. It is what is measured 
in practice to determine the ratio $n$($^6$Li)/$n$($^7$Li). 

For given $T_{\rm eff}$ and $\log g$, the predicted line asymmetry is 
fairly constant for metallicities below [Fe/H]=$-2$, but increases 
noticeably towards higher [Fe/H]. Hotter stars with lower gravity 
(turn-off stars) show higher line asymmetries than cooler unevolved 
dwarfs.

 \begin{table}
 \centering
 \begin{tabular}{llllll}
\hline 
 $T_{\rm eff}$ & $\log g$ & [Fe/H] & asym. & $N$ & $X\times Y \times Z$ [Mm$^3$] \\
\hline
6250 & 4.0 & -3.0    & 0.020    & 20 & $26\times 26\times 12.7$\\            
6250 & 4.0  & -2.0   &  0.021   & 16 & $26\times 26\times 12.7$\\  
6250  & 4.0 &  -1.0  &  0.037   & 20 & $26\times 26\times 12.7$\\ 
6250 & 4.5  &  -3.0  &   0.012  & 18 & $7.0\times 7.0\times 3.9$\\
6250 & 4.5 & -2.0   & 0.011     & 19 & $7.0\times 7.0\times 3.9$\\
6250 & 4.5 &  -1.0  & 0.017     & 20 & $7.0\times 7.0\times 3.9$\\
\hline
5900 & 4.0 & -2.0 &  0.015      & 20 & $26\times 26\times 12.5$\\
5900 & 4.5  & -2.0 & 0.011      & 19 & $6.0\times 6.0\times 3.8$\\
\end{tabular}
\caption{Computed line asymmetry expressed as equivalent isotopic ratio
  $^6$Li/$^7$Li (asym.), and 3D model properties: $T_{\rm eff}$ is the 
  effective temperature, $\log g$ the surface gravity, [Fe/H] the 
  metallicity, $N$ the
  number of considered statistically independent snapshots, 
  and $X\times Y \times Z$ the geometrical extend of the computational box. 
  All models use six wavelength bins to represent the
  wavelength dependence of the radiative energy exchange, and a grid of
  $140\times 140 \times 150$ cells.}
\label{assymprofile1}
\end{table}

\subsection{The source of the line asymmetry}
It is interesting to have a look at the dynamical origin of the line asymmetry.
Obviously, it is the hydrodynamical asymmetry between upward and
downward convective flows that is at the origin of the asymmetry 
of the 3D line profiles.
Fig.~\ref{z-motion} shows a histogram of the vertical velocity at 4
optical depths in a typical 3D hydrodynamical model ($T_{\rm eff}=6250$~K, 
$\log g=4.0$, [Fe/H]$=-2.0$). There is little
asymmetry at $\log \tau_\mathrm{ross}=-3$ or $-2$, but in the lower
photosphere ($\log \tau_\mathrm{ross}=-1$ and $0$) the asymmetry is
conspicuous.
For the centre of the disc, the 3D-NLTE contribution function of the Li\,I 
resonance line peaks at $\log \tau_\mathrm{ross} \approx -1$. 
The broadening 
is large and the asymmetry is important (Fig.~\ref{z-motion}). In 3D-LTE,
however, the contribution function has a double peak, one at
$\log \tau_\mathrm{ross} \approx -1$, but a second one at in very 
superficial and very cool layers \cite{CSC07}, where the velocity distribution
is narrower and virtually symmetric.
This explains why the 3D-LTE analysis of ALN06 differs from our 3D-NLTE
analysis. There is a substantial difference in the two profiles, the 3D-LTE
profile being a superposition of an asymmetric line (the part formed in the
lower photosphere) and of a symmetric line (with a different shift)
formed in the higher photospheric layers.

Clearly, the real situation is more complex than suggested by the simplistic
picture outlined above. It is not only the distribution of the vertical velocity 
but also its correlation with the temperature fluctuations that determines
the shape of the line profile. Moreover, horizontal motions also become 
important when integrating the line profile across the stellar disk.

\subsection{Conclusion}
The level of the plateau attributed to\ $^6$Li in ALN06 \cite{ALN06} is 
affected by a bias which, when corrected, moves the plateau to a mean 
value of\ $^6$Li/$^7$Li$\approx 0.022$. The study
of the\ $^7$Li line asymmetry with a 3D-NLTE code on 3D hydrodynamical models 
obtained with the CO$^5$BOLD code leads to asymmetry signals of respectively 
0.020, 0.021 for [Fe/H]$=-3.0$ and $-2.0$ respectively, at $T_{\rm eff}=6250$~K, 
$\log g=4.0$.
Stars corresponding to the subsample of ALN06 with $6100$~K$<T_{\rm eff}<6400$~K,
$ 3.7<\log g<4.3$,have a ratio $n$($^6$Li)/$n$($^7$Li) with a mean value of 0.021 (14 objects).
For metallicity [Fe/H]=$-1$, the predicted asymmetry signal is 0.037, and the observed 
value in ALN06 is 0.023, not a very good fit but the number of objects is only 4.
The effect of the gravity is a decrease of the asymmetry by a factor of 2 when 
$\log g=4.5$ instead of $4.0$. The only certain unevolved star in the
ALN06 sample is HD19445: observed signal: 0.002, asymmetry predicted 0.010.     
The conclusion is that the  plateau of the remarkable ALN06 set of 
observations can be reinterpreted without a contribution by\ $^6$Li.
However, a detailed re-analysis of a few stars with the highest apparent
$^6$Li content is necessary before a final conclusion can be drawn for them.

\end{document}